\documentclass[letterpaper,twocolumn,10pt]{article}

\usepackage[usenames,dvipsnames]{xcolor}

\usepackage{amsmath,amssymb,amsthm}
\usepackage{mathptmx}

\usepackage{xspace}
\def\ie{{i.e.\xspace}}
\def\eg{{e.g.\xspace}}
\def\etal{{et al.\xspace}}

\def\etc{etc.\xspace}

\usepackage{mathtools}

\DeclareMathOperator*{\E}{\mathbb{E}}
\newcommand{\diff}{\mathop{}\!\mathrm{d}}

\usepackage{makecell}
\usepackage{booktabs}
\usepackage{multirow}

\usepackage{subcaption}

\usepackage{tikz}
\newcommand*\circled[1]{%
\begin{tikzpicture}[baseline=(C.base)]
    \node[shape=circle,draw,inner sep=1pt](C) {#1};
\end{tikzpicture}%
}

\graphicspath{{figures/}}

\usepackage[small, compact]{titlesec}
\usepackage[textfont={it}, font={small,bf}, skip=1pt]{caption}

\usepackage[inline]{enumitem}
\newlength{\docparskip}
\setlist[enumerate]{
    nosep,
    topsep=2pt,
    partopsep=4pt,
    itemsep=2pt,
    leftmargin=1.5em,
}
\setlist[itemize]{
    nosep,
    topsep=2pt,
    partopsep=4pt,
    itemsep=2pt,
    leftmargin=1.5em,
}
\setlist[description]{
    nosep,
    topsep=2pt,
    partopsep=4pt,
    itemsep=2pt,
}
\newlist{inlineenum}{enumerate*}{1}
\setlist[inlineenum]{label=\itshape\alph*\upshape)} 

\usepackage[
  backend=biber,
  style=numeric-comp,
  maxbibnames=99, 
  minbibnames=10, 
  sortcites=true,
  hyperref=true,
]{biblatex}
\usepackage[
  ruled,
  vlined,
  nofillcomment,
  linesnumbered,
]{algorithm2e}
\DontPrintSemicolon{}
\SetKwComment{Comment}{$\triangleright$\ }{}

\usepackage{usenix-2020-09} 

\hypersetup{
  colorlinks,
  linkcolor={red!70!black},
  citecolor={red!70!black},
  urlcolor={blue!70!black}
}

\makeatletter
\let\oldtheequation\theequation 
\renewcommand\tagform@[1]{\maketag@@@{\ignorespaces#1\unskip\@@italiccorr}}
\renewcommand\theequation{(\oldtheequation)}
\makeatother

\usepackage{placeins}

\newcommand{\name}{\textsc{Orloj}\xspace}

\begin{document}

\date{}

\title{\Large \bf \name: Predictably Serving Unpredictable DNNs}

\author{
  {\rm Peifeng Yu}\\
  peifeng@umich.edu\\
  University of Michigan
  \and
  {\rm Yuqing Qiu}\\
  qyuqing@umich.edu\\
  University of Michigan
  \and
  {\rm Xin Jin}\\
  xinjinpku@pku.edu.cn\\
  Peking University
  \and
  {\rm Mosharaf Chowdhury}\\
  mosharaf@umich.edu\\
  University of Michigan
} 

\maketitle

\begin{abstract}
Existing DNN serving solutions can provide tight latency SLOs while maintaining high throughput via careful scheduling of incoming requests, whose execution times are assumed to be highly predictable and data-independent.
However, inference requests to emerging dynamic DNNs -- \eg, popular natural language processing (NLP) models and computer vision (CV) models that skip layers -- are \emph{data-dependent}.
They exhibit poor performance when served using existing solutions because they experience large variance in request execution times depending on the input -- the longest request in a batch inflates the execution times of the smaller ones, causing SLO misses in the absence of careful batching.

In this paper, we present \name{}, a dynamic DNN serving system, that captures this variance in dynamic DNNs using empirical distributions of expected request execution times, and then efficiently batches and schedules them without knowing a request's precise execution time.
\name{} significantly outperforms state-of-the-art serving solutions for high variance dynamic DNN workloads by 51--80\% in finish rate under tight SLO constraints, and over 100\% under more relaxed SLO settings.
For well-studied static DNN workloads, \name{} keeps comparable performance with the state-of-the-art.

\end{abstract}

\section{Introduction}%
\label{sec:intro}

Deep neural network (DNN) model inference requests constitute an increasingly larger portion of today's web requests~\cite{hazelwood2018applied,wu2019machine}.
For example, NVIDIA estimated that 80--90\% of the cloud AI workload was inference processing~\cite{leo19}. 
It is only likely to increase for the foreseeable future with the proliferation of DNN-powered APIs~\cite{google-document-ai,github-copilot} that enable applications to build on top of pre-built foundation models~\cite{bommasani2021opportunities}.

The underlying serving systems~\cite{clipper,tfserving,infaas,nexus,clockwork} that handle these inference requests aim to maximize throughput while reducing service level objective (SLO) misses.
Due to the high throughput and low latency requirements of DNN-dependent applications and the large computation needs of DNN inference requests, modern serving systems often rely on expensive GPUs to serve many requests in parallel by batching them together~\cite{clockwork}.
All the requests in the same batch experience the same request execution time.
This works well because the state-of-the-art DNN serving systems~\cite{clipper,nexus,clockwork} have so far assumed \emph{data-independence} of incoming requests; \ie, the amount of computation required for each request is the same regardless of the input data.
For example, for an image classification model, whether the input image contains a dog or a cat, the model performs exactly the same computation to derive the answer.
Consequently, the request execution time of a model can be accurately profiled and used to precisely schedule inference requests~\cite{clockwork}.

We observe that the recent rise of a new class of \emph{dynamic} DNNs~\cite{transformer,dnn-survey} challenges this assumption (\autoref{sec:ddnn}).
Unlike static DNNs, dynamic DNNs can adapt their structures or parameters to the input during inference and are thus inherently \emph{data-dependent}.
For example, SkipNet~\cite{skipnet} dynamically skips layers depending on the input sample;
RDI-Nets~\cite{rdi-nets} allows for each input to adaptively choose one of the multiple output layers to output its prediction;
and various NLP models exhibit recurrent structures or loops~\cite{gpt,bart,fsmt,mbart,t5,blenderbot}.
The result is unpredictable execution times for individual requests.
Because request execution times come from a distribution instead of being a single constant, existing systems that use a single mean or tail execution time from historical data to \emph{plan ahead} perform poorly~\cite{clockwork, nexus}.
They fail to capture the high variance in incoming requests' execution times, and when they batch multiple requests together, one long request in the batch slows down many short ones, leading to large number of SLO misses (\autoref{sec:motivation}).
Serving systems that do not assume data independence~\cite{clipper,inferline,infaas} and instead perform \emph{reactive} adjustment to dynamically provision workers at runtime perform even worse, because they treat SLOs as long-term reactive targets and cannot effectively curtail tail latency, especially under stringent SLO constrains~\cite{clockwork}.

In this paper, we present \name, a distribution-aware dynamic DNN serving system, to provide high throughput and low SLO misses.
\name{} also takes a plan-ahead approach, but unlike recent solutions, it uses a random variable to capture the variance in request execution times of dynamic DNNs, rather than assuming a constant mean or tail latency for all requests.
This gives \name{} more flexibility to account for uncertainty in execution times when batching them together to achieve high throughput.

Going beyond prior distribution-based solutions for cluster scheduling and query processing~\cite{chiDistributionbasedQueryScheduling2013, 3sigma,tiresias:nsdi19}, \name{} addresses two challenges unique to model serving.
First, batching is vital to achieve high throughput, but it also affects execution times, as all requests in the same batch start and finish at the same time, regardless of their individual execution times.
\name{} proposes a batch-aware priority score that derives batch execution time distributions given those of individual requests, and uses the score to guide its batching decisions.
%
Second, because a model can receive requests from different applications, the joint distribution can be multimodal with even higher variance.
To this end, \name{} tags each request with its originating application and relies on probability theory to accurately estimate batch execution times even when execution times of requests in the same batch follow different distributions.

We have implemented and evaluated \name{} using production traces and a large range of possible input execution time distributions (\autoref{sec:eval}).
In comparison to Clockwork~\cite{clockwork}, Nexus~\cite{nexus}, and Clipper~\cite{clipper}, \name{} can improve the finish rate when serving dynamic DNNs by 51--80\% under tight SLO constraints, and over 100\% under more relaxed SLO settings.
For well-studied static DNN workloads, \name{} keeps comparable performance with the state-of-the-art.

Overall, we make the following contributions in this paper:
\begin{itemize}
    \item To the best of our knowledge, \name{} is the first system to systematically analyze the inference performance of dynamic DNNs and associated challenges.

    \item We present a batch-aware distribution-based scheduling algorithm to handle batching and multimodal distributions in dynamic DNN serving systems.

    \item \name{} can handle both static and dynamic workloads with high throughput and tight SLO guarantees.
\end{itemize}

\section{Background and Motivation}

In this section, we overview model serving systems and the recent rise of dynamic DNNs, and then move on to the limitations of existing state-of-the-art solutions when serving such dynamic networks.

\subsection{Model Serving}%
\label{sec:mlserving}

\begin{figure}
    \centering
    \includegraphics{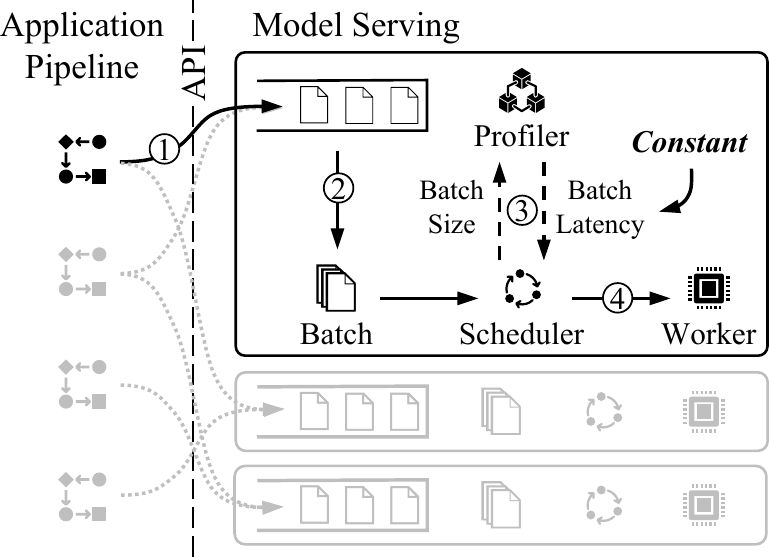}
    \caption{A model serving service multiplexes requests from multiple applications and schedules batches of requests on workers.}%
    \label{fig:pipeline-worker}
\end{figure}

Increasingly more models are deployed on the critical paths of online interactive services~\cite{wu2019machine}.
They may even comprise dependent computations across multiple models, forming pipelines~\cite{inferline}.
For example, video analytics pipelines first detect certain objects from each image and then recognize individual objects~\cite{video-pipeline, nexus}.
A serving system in the backend handles inference requests for individual models by running a model replica on the input,
usually providing APIs for specific tasks such as detection, translation, or prediction~\cite{google-document-ai,github-copilot}.
Similar to other datacenter services~\cite{adya2016slicer}, it multiplexes workloads of different applications and load balances requests across multiple workers~\cite{infaas, clipper}.


\paragraph{Lifecycle of Serving Inference Requests}
\autoref{fig:pipeline-worker} illustrates the lifecycle of (a batch of) inference requests in a worker of such serving systems.
\circled{1} Incoming requests first go through a priority queue, usually ordered by deadline, but it may vary depending on a system's optimization goal.
\circled{2} The dynamic batcher will extract requests from the queue to create batches, while respecting the deadline requirement of the requests at the top of the queue.
\circled{3} The size of the formed batch will be queried against historical profiling data to decide an estimated batch latency.
\circled{4} This single latency value will be used in the scheduler to derive an execution plan on the worker.
Of course, there is not exactly one way to divide work between the stages shown here, and there may be additional interactions between various components.
For example, Nexus~\cite{nexus} uses a pre-computed plan ahead of time,
while Clockwork~\cite{clockwork} employs multiple ``Batch Queues'' to select the best batch size at runtime.
Nevertheless, the active state in these systems depends on a \emph{single, point estimation of the batch latency}, oblivious to individual request-specific data at runtime.

\paragraph{Batching and the Throughput-Latency Tradeoff}
To achieve high throughput, model serving systems~\cite{clockwork,nexus,clipper,inferline,infaas} rely on batching multiple requests.
Requests arriving within a specific time window are batched together to ensure that worker resources do not idle~\cite{clipper,nexus}.
There is, however, a tradeoff in picking a batch size.
While a larger batch size may increase throughput, it also means longer time window; the latter can lead to higher (tail) latency and missed SLOs.
In contrast, a smaller batch size can result in underutilized hardware.
Existing solutions focus on finding the sweet spot to reduce SLO misses, typically for each model individually~\cite{clockwork,nexus,clipper,infaas},
while pipeline-aware solutions break down the end-to-end SLO into smaller pieces~\cite{inferline}.
The overall objective of a model serving system can, therefore, be described as \emph{maximize throughput while reducing SLO misses}.

\subsection{Dynamic DNNs}%
\label{sec:ddnn}

\begin{figure}[t]
    \centering
    \subcaptionbox{RDI-Nets}{
        \includegraphics{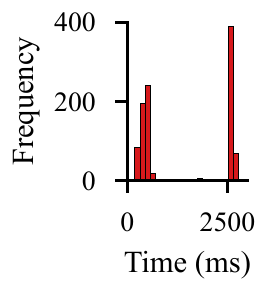}
    }
    \subcaptionbox{GPT}{
        \includegraphics{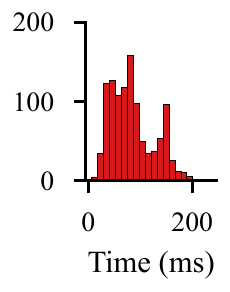}
    }
    \subcaptionbox{Inception V3}{
        \includegraphics{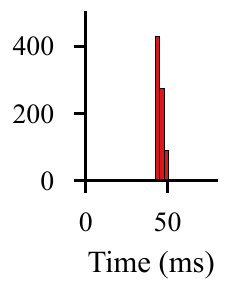}
    }
    \subcaptionbox{SkipNet}{
        \includegraphics{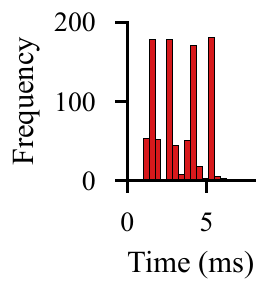}
    }
    \subcaptionbox{BART}{
        \includegraphics{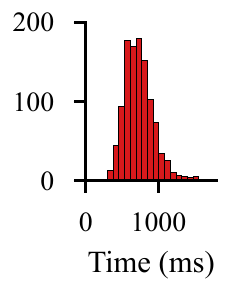}
    }
    \subcaptionbox{ResNet}{
        \includegraphics{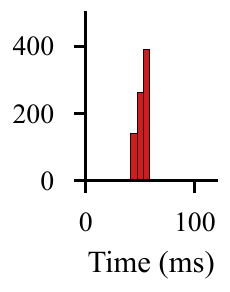}
    }
    \caption{Inference request execution time varies widely in dynamic CV and NLP models. (Inception V3 and ResNet are shown for comparison.)}%
    \label{fig:infer-latencies}
\end{figure}

Point estimations in existing solutions is sufficient only because they focus on static DNNs (e.g., CNNs), where each request roughly takes the same amount of time and is highly predictable~\cite{clockwork}.
However, we observe that dynamic DNNs are becoming increasingly more popular in recent years~\cite{bommasani2021opportunities, dnn-survey}.
Dynamic DNNs adapt their structures or parameters to different inputs, leading to notable advantages in terms of accuracy, computational efficiency, adaptiveness, \etc, compared to static DNNs that have fixed computational graphs and parameters at the inference stage~\cite{dnn-survey}.
Examples of dynamic DNNs include various language models~\cite{bert,transformer} as well as early-exit techniques used in emerging CNN models~\cite{skipnet}.

\paragraph{Observation: Dynamic DNN Inference is Unpredictable}
As the name suggests, the computation requirement of inference requests to a dynamic DNN model can be \emph{dynamic}.
It naturally follows that the request execution time is no longer constant for different inputs anymore.
We report inference execution time histograms for four common dynamic networks (SkipNet~\cite{skipnet}, RDI-Nets~\cite{rdi-nets} for image recognition, GPT~\cite{gpt}, BART~\cite{bart} for NLP) in \autoref{fig:infer-latencies}.
For comparison, we also include the execution time for two common static CV models: Inception V3~\cite{inception} and ResNet~\cite{resnet}.
Note that the absolute number of requests in these histograms is less important, as it merely represents the testing dataset we used.
However, the existence of a large range of values in the x-axis reflects the possibility of different execution times.

For the image recognition models, there are a few distinct clustered ranges, which represent multiple code paths with different execution times, created by skipping/choosing parts of the model.
Similar observations hold for NLP models too. 
However, instead of having distinct code paths, execution time for NLP models falls in a continuous range, reflecting the impact of the input sequence length on execution time.
Nevertheless, it can be seen that the difference in inference latency can be as large as $10\times$, with some requests finishing in $10$ms while many others taking more than $100$ms.

\begin{figure}[!t]
    \centering
    \includegraphics{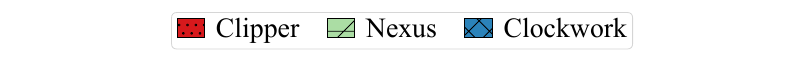}
    \subcaptionbox{
        Simple normal distribution%
        \label{fig:existing-finish-rate-incoming-one_modal}%
    }{
        \includegraphics{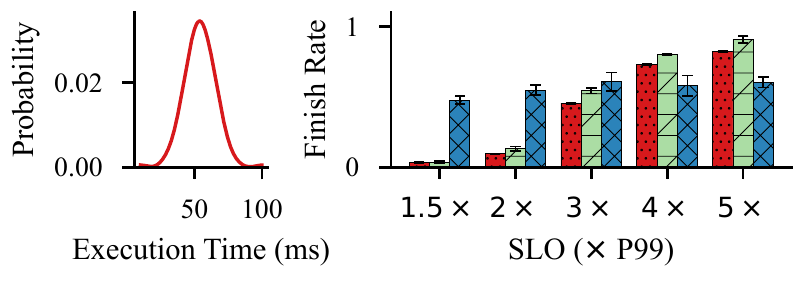}
    }
    \subcaptionbox{
        Bimodal distribution%
        \label{fig:existing-finish-rate-incoming-two_modal}%
    }{
        \includegraphics{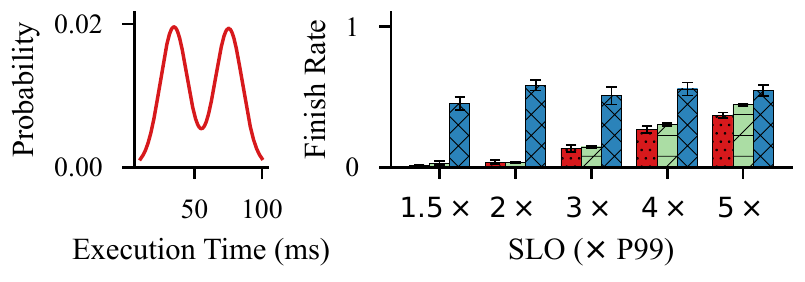}
    }
    \subcaptionbox{
        Bimodal distribution w/ inequal peaks%
        \label{fig:existing-finish-rate-incoming-not_equal}%
    }{
        \includegraphics{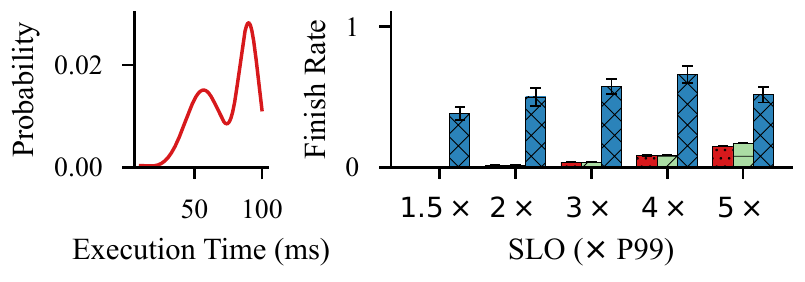}
    }
    \caption{Performance of existing model serving solutions for dynamic DNNs. Similar results hold for more diverse distributions as well (\autoref{sec:eval}).}%
    \label{fig:existing-finish-rate}
\end{figure}

\paragraph{Challenges in Serving Dynamic DNNs}%
\label{sec:challenges}
The presence of a large variance in execution times of dynamic DNNs as well as the shared nature of model-serving-as-a-service systems lead to two key challenges when serving dynamic DNNs.

\begin{enumerate}
    \item \textbf{Effective batching:}
    Batching is a must for high throughput and worker utilization.
    However, all requests in the same batch start and finish at the same time, regardless of their individual execution times.
    When requests in the same batch vary widely in their execution times, the longest one becomes the straggler and slows down the entire batch.

    \item \textbf{Handling multimodal distribution:} A model built for a specific task (\eg, classification, translation, etc.), is often used by multiple applications, especially when it is exposed as a service.
    As a result, input requests and their corresponding execution times often follow different application-specific distributions.
    The combined multimodal distribution has even higher variance, which can introduce even more stragglers during batching.
\end{enumerate}

\subsection{Limitations of Existing Solutions}%
\label{sec:motivation}

Indeed, state-of-the-art model serving solutions~\cite{nexus,clipper,clockwork}, which have been optimized for static DNNs, suffer from high SLO violation rates when applied to dynamic DNNs.
\autoref{fig:existing-finish-rate} shows the finish rates of three recent model serving systems when the input execution time follows various distributions, under different SLO settings.
For each case, the probability distribution function (PDF) of the input execution times is shown to the left.
For all cases,
the incoming rate trace is derived from the Microsoft Azure Functions workload trace~\cite{azure} similar to Clockwork~\cite{clockwork}.
\autoref{sec:eval} provides more details on the methodology.

The high-level takeaway is that all these systems have undesirable performance.
As most batches contain both long requests and short ones, the execution time for the whole batch is almost always longer than the average.
This causes Clockwork~\cite{clockwork} to often mispredict a batch's latency, which in turn leads to frequent time-out error in its scheduler, causing the subsequent batch to fail.
This explains its close-to-half finish rate.
Nexus~\cite{nexus} pre-computes an execution plan ahead of time using the average execution time, but due to the variance in input execution, it cannot reach a stable state.
Clipper~\cite{clipper} monitors request execution time reactively, but it cannot keep up under tight SLO settings.
Fundamentally, we observe the effect of existing solutions failing to batch requests effectively.

\paragraph{Distribution-Based Schedulers}
Existing distribution-based schedulers such as 3Sigma~\cite{3sigma} or Shepherd score~\cite{chiDistributionbasedQueryScheduling2013}, proposed for cluster scheduling and query processing respectively, do not fare well either. 
They do not consider sub-second level latency constraints or inference serving-specific challenges like batching and lack of preemption.

\section{\name{} Overview}%
\label{sec:overview}

{\name} is an inference serving system that serves inference requests to a dynamic DNN model while maximizing the number of requests that can be served within the SLO\@.
In this section, we provide an overview of how {\name} fits in the inference life cycle of dynamic DNNs to help the reader follow the subsequent sections.

\subsection{Problem Statement}

Each inference request in {\name} is defined by its \emph{release time} and \emph{deadline} (release time plus SLO), and has a minimum \emph{execution time} that is measured when the request is executed alone.
Multiple applications with diverse use cases and corresponding input distributions send requests to the same dynamic DNN model served by {\name}.
Each GPU worker processes these requests one batch at a time.
Note that, to scale out to a pool of workers in a cluster setting, different models and their replicas can use {\name} in parallel.

Given a set of pending inference requests, the {\name} scheduler must decide \emph{which subset of them should be included in the next batch} submitted to the GPU to maximize throughput while reducing SLO misses, under the following constraints:

\begin{itemize}
    \item \textbf{Partial information}: the execution time of a request, and thus a batch, is unknown to the scheduler. 
    However, its probabilistic distribution can be learned from historical data.

    \item \textbf{Non-preemption}: inference execution of a batch cannot be preempted after it is submitted to the GPU\@.
\end{itemize}

\subsection{\name{} Architecture}

Unlike existing model serving systems, \name{} represents the request execution time of a dynamic DNN model as a random variable, which is described using an empirical distribution over a time window.
Thereafter, instead of using simply the mean or the max of the population, it makes scheduling decisions using the knowledge of the entire distribution.

\begin{figure}
    \centering
    \includegraphics{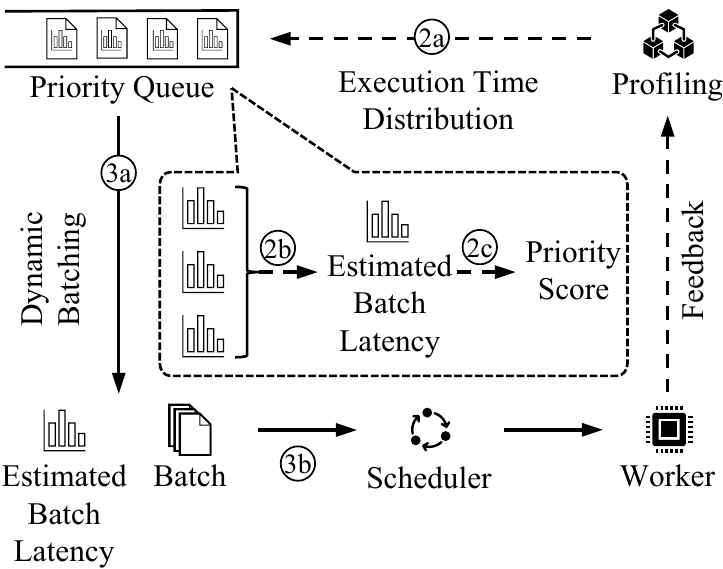}
    \caption{{\name} architecture.}%
    \label{fig:architecture}
\end{figure}

\name{} addresses the challenges arising from multimodal distribution and effective batching by proposing a \emph{time-varying priority score} (detailed in \autoref{sec:algo}) by considering a batch's combined distribution.
It determines this priority score for all requests that potentially can be put together to form a batch, and then it performs priority-based scheduling.

\paragraph{Inference Lifecycle}

As shown in \autoref{fig:architecture}, \name{} follows the same overall process as other model serving systems, but with updated scheduling steps \circled{2} and \circled{3} in \autoref{fig:pipeline-worker}:
\begin{enumerate}
    \item[\circled{2a}] incoming requests are tagged per application, and the application-specific execution time distribution is associated with each request using the information collected by {\name}'s online profiler;
    \item[\circled{2b}] execution time distributions from requests are combined to derive estimated batch execution time (latency);
    \item[\circled{2c}] next, the priority score for all requests are calculated using the estimated request latency, deadlines and the current time as input;
    \item[\circled{3a}] {\name} calculates estimated batch latency for all potential batch sizes;
    \item[\circled{3b}] the scheduler loop selects a feasible batch size to actually create the batch.
\end{enumerate}
After obtaining a batch, \name{} submits it to the worker for execution, following the same step \circled{4} as existing model serving systems.

Next, we elaborate on \name{}'s scheduling loop and its batch handling.

\paragraph{Batch Size Selection}
It is not always possible to execute requests using the maximum possible batch size, because some requests may have tighter deadlines than others and waiting for the maximum batch size can take too long.

\name{} tracks the set of feasible batch sizes for each request.
These sets of feasible batch sizes are updated over time.
When the deadline is approaching, batch sizes that are too large to meet the deadline will be dropped, so the corresponding request will only be considered for small batches.
In addition, for each batch size, \name{} keeps note of the earliest deadline for requests suitable for the batch size.
The batch size with the overall earliest deadline will be chosen by the scheduler to lazily create a batch and send to the worker for execution.

\autoref{alg:dyninfer-iter} shows \name{}'s scheduling loop, which implements the above batch size selection scheme (\autoref{algline:start-drop} to \autoref{algline:end-batch}).
It is worth noting that \name{} uses a separate priority queue $Q_{\mathit{bs}}$ for each batch size $\mathit{bs}$.
The earliest deadline for requests in $Q_{\mathit{bs}}$ is tracked by an additional Fibonacci heap to allow online deletion.

\paragraph{Batch Priority}
The highest priority batch is the one to be scheduled next, and it depends on the priorities of individual requests.
\name{} uses a time-varying score for request priority, \ie, the priority of a request changes over time.
Naively sorting the pending requests' queue in $O(n \log n)$ time for each scheduling iteration in the hot path is inefficient.
Instead, we use an $O(\log^2 n)$ priority queue~\cite{icbs,chiDistributionbasedQueryScheduling2013}.
We also address issues related to floating-point overflow when using such queue in practice (\autoref{sec:convex-hull}).

\begin{algorithm}[!t]
    \linespread{1}\selectfont

    \caption{\name{} Scheduler Iteration}\label{alg:dyninfer-iter}

    \SetKwFunction{milestone}{Milestone}
    \SetKwFunction{EstimateBatchLatency}{EstimateBatchLatency}
    \SetKwFunction{PopBatch}{PopBatch}

    \KwIn{\\
        \Indp{}
        $\mathcal{R} \leftarrow $ set of pending requests, \\
        $t \leftarrow$ current time, \\
        $\mathcal{S} \leftarrow$ set of batch sizes suported by the model, \\
        $Q_{\mathit{bs}} \leftarrow$ set of requests viable for batch size $\mathit{bs}$, \\
        $D_r \leftarrow$ deadline of request $r$, \\
        $D_{Q_{\mathit{bs}}} \leftarrow$ $\min \{D_r | r \in Q_{\mathit{bs}}\}$, \\
    }
    \KwOut{Batch $\mathcal{B} \subseteq \mathcal{R}$}

    \Comment{Update priority scores (\autoref{sec:convex-hull})}

    $U \leftarrow \emptyset$ \Comment*{requests needs to be updated}\label{algline:start-update}

    \If{need reset base time}{
        reset base time \;
        $U \leftarrow \mathcal{R}$ \;
    }
    \For{$r \in \mathcal{R}$}{
        \If{$t \ge \milestone{r}$}{
            $U \leftarrow U \cup \{r\}$
        }
    }
    \For{$r \in U$, $\mathit{bs} \in \mathcal{S}$}{
        update priority score of $r$ in $Q_{\mathrm{bs}}$
    }\label{algline:end-update}

    \Comment{Drop requests from queue if too late}

    \For{
        $\mathit{bs} \in \mathcal{S}$, $r \in Q_{\mathit{bs}}$ ordered by $D_r$%
        \label{algline:start-drop}
    }{
        \If{$t + \EstimateBatchLatency{r, bs} > D_r$}{
            $Q_{\mathit{bs}} \leftarrow Q_{\mathit{bs}} \setminus \{r\}$ \;
        }

        \If{$\mathit{bs}$ is the last feasible batch size for $r$}{
            Mark $r$ as timed out \;
        }
    }\label{algline:end-drop}

    \Comment{Determine candidate batch size}
    $\mathit{candidate} \leftarrow \mathsf{nil}$ \;\label{algline:start-batch}
    \For{$\mathit{bs}$ ordered by ($D_{Q_{\mathit{bs}}}$, $\mathit{bs}$) in descending order} {
        \If{$\lvert Q_{\mathit{bs}} \rvert \ge \mathit{bs}$} {
            $\mathit{candidate} \leftarrow \mathit{bs}$ \;
            break \;
        }
    }
    \If{$\mathit{candidate}$ is $\mathsf{nil}$} {
        return \;
    }
    \Comment{Select top ones ordered by \name{} score}
    return $\mathcal{B} \leftarrow$ \PopBatch{$Q_\mathit{candidate}$} \;\label{algline:end-batch}
\end{algorithm}

Before proceeding to the details of \name{}'s algorithm in the next section, we highlight a few other components in \name{}.

\paragraph{Per-Application Tracking}
As shown in \circled{2a}, \name{} associates each request an execution time distribution using application-specific historical data.
First, it is possible to distinguish requests from application as there are usually certain application IDs involved when the model is exposed as a service.
Second, such tracking is also necessary, because applications may solve problems in different domains despite using the model for the same task.
As a result, input requests execution times often follow different distributions.
While \name{} does not assume any pre-defined distribution for its input and only tracks empirical distributions, the combined multimodal distribution has higher variance. 
This hurts scheduling abilities even if the scheduler has perfect information of its distribution, because the scheduler has to account for different possibilities when scheduling.

\paragraph{Long-Term Feedback Loop}%
\label{sec:auto}
Incoming requests can change their arrival pattern and volume over time, either due to the diurnal nature of the service or due to shifts in general interests.
\name{} therefore needs to track per application execution time data for requests over time.
However, our calculation needs the execution time for requests when they execute alone, which cannot be guaranteed if simply measuring the time online.
Instead, the profiler in \name{} takes an asynchronous approach.
Finished requests are sampled and send to the profiler to evaluate individually.
The execution time data will then be asynchronously picked up and accumulated by the scheduler periodically, completely off the critical path.
In order to adapt to drifts in the input, \name{} resets its profiling memory every once a while.
The exact window is configurable and is determined by domain knowledge.

\section{Batch-Aware Distribution-Based Scheduling}%
\label{sec:algo}

At its core, \name{} is a priority-based scheduler where the priorities of individual requests are determined using a cost function that captures the distribution of request execution times.
Highest priority requests are then put in a batch to achieve the maximum level of parallelism, which is then submitted to the worker.
To achieve this, \name{} relies on probability theory to accurately estimate request execution times even when requests in the same batch affect each other and their executions are no longer independent.

In this section, after a brief introduction of cost function and the definition of priority on a single request (\autoref{sec:algo-prelim}), we dive into the derivation of a vital term in the batch-aware priority score -- batch execution time,
given request execution time following the same distribution (\autoref{sec:algo-same}) and different distributions (\autoref{sec:algo-diff}).
Finally, we discuss how to break the cyclic dependency between batch formation and priority score computation (\autoref{sec:batch-formation}), as well as floating-point overflow handling when applying the algorithm in practice (\autoref{sec:convex-hull}).

\subsection{Preliminaries}%
\label{sec:algo-prelim}

\begin{figure}[t]
    \centering
    \includegraphics{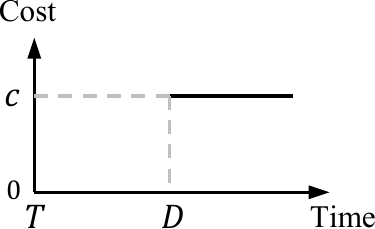}
    \caption{An example of SLO cost function.}\label{fig:cost-fn}
\end{figure}

\paragraph{Cost Function}
Instead of directly optimizing for metrics such as average/tail latency and throughput, we model SLO deadlines using a cost function that captures the opportunity cost differences between two important scheduling decisions.
On the one hand, executing a request involves costs for resource usage (\eg, server utilization) and opportunity cost (\eg, it may postpone other requests).
On the other hand, missing deadline may have a monetary penalty according to the SLO\@.
Throughout the rest of the paper, we use SLO cost functions similar to that in \autoref{fig:cost-fn}.
Meaning, for requests arriving at time $T$ with deadline $D$, there is a penalty $c$ for missing that deadline.

\paragraph{Scheduler Objective}
The objective of our scheduler is to minimize the overall cost, or as we set out to do, maximize the number of requests that finish within corresponding deadlines.
%
Selecting a request to put into the next batch reduces the expected cost that would have been incurred if it were delayed.
Therefore, our goal is to find requests for which the ratio of expected cost reductions are the greatest.

\paragraph{Background: Priority of a Single Request}
Consider a request whose execution time $L$ is a random variable and its cost function is $C(t)$.
Cost reduction for this request boils down to the difference between the costs of two scheduler decisions: $C_{\mathit{now}}(t)$, including the request in the next batch and executing it right away, or $C_{\mathit{delay}}(t)$, selecting another one and thus delaying this request.

Its priority $p(t)$ can thus be defined as:

\begin{equation}\label{eq:pt-orig}
p(t) = \frac{1}{\E \lbrack L \rbrack} \left(
    \E \lbrack
        C_{\mathit{delay}}(t)
    \rbrack
    - \E \lbrack C_{\mathit{now}}(t) \rbrack
\right)
\end{equation}

Note that $C_{\mathit{now}}(t) = C(t + L) $ is a random variable, as well as $C_{\mathit{delay}}(t) = C(t + \tau + L)$ ($\tau$ is the anticipated delay).

Given that $C(t)$ has the same shape as in \autoref{fig:cost-fn}, and assuming $\tau$ follow an exponential distribution with parameter $b$,\footnote{The probability that a request will be selected for execution, given that it is already queued, does not change with time unless there is a change in the state of the queue. Therefore, the anticipated delay is an exponential~\cite{cbs}.} prior work~\cite{chiDistributionbasedQueryScheduling2013} has shown that, when $L$ can be described using a histogram, $p(t)$ can be derived by computing on each histogram bin separately and combining the results:

\begin{equation}\label{eq:pt}
\begin{aligned}
    p(t) &= \sum_{i} p_i(t) \\
    p_i(t) &=
        \begin{cases}
            \frac{hc}{\E\lbrack L\rbrack b} \left( e^{bl^{(i)}_2} - e^{bl^{(i)}_1} \right) e^{-bD} e^{bt}
            & t < D - l^{(i)}_2 \\
            \frac{hc}{\E\lbrack L\rbrack b} - \frac{hc}{\E\lbrack L\rbrack b} e^{bl_1} e^{-bD}e^{bt} & D - l^{(i)}_2 \leq t < D - l^{(i)}_1 \\
            0 & D - l^{(i)}_1 \leq t
        \end{cases}
\end{aligned}
\end{equation}
where $p_i(t)$ is the score for the $i$-th bin in the histogram with range $[l^{(i)}_1, l^{(i)}_2)$ and frequency $h$. 
We represent the deadline for the request in consideration using $D$, with $c$ being the cost when missing the deadline.

\subsection{Batch Latency Estimation}%
\label{sec:batching}
We still need to find out the distribution of $L$, as well as $\E \lbrack L \rbrack$.
As we discussed in \autoref{sec:mlserving}, the scheduler must schedule \emph{batches} of requests to ensure high throughput.
This is an important detail, because up until now, we assumed the execution time $L$ as an intrinsic property of the request, depending solely on the request itself.
However, under the batch execution model, it is no longer the case: \emph{requests do not execute alone, and they affect each other's execution time in the same batch.}
The purpose of the term $\frac{1}{\E\lbrack L \rbrack}$ in \autoref{eq:pt-orig} is to account for the worker time usage of the request.
As such, to accurately represent a request's potential worker time usage during batching, $L$ must now be the execution time of the whole batch the request is in.

When the request execution time itself is a constant (\ie, in static DNN scenarios), this is trivial: requests are homogeneous; so is the batch.
It is thus possible to profile the batch execution time for all possible batch sizes ahead of time~\cite{clockwork}.
In our case, however, not only are requests' execution times random variables, but \emph{a batch may also contain requests from different duration distributions.}
Next, we describe how \name{} handles batches of requests following the same or different distribution separately.

\subsubsection{Requests in Batch Follow the Same Distribution}%
\label{sec:algo-same}
For a batch $B$ of $k$ requests, if requests are of the same duration $l$, the batch execution time $l_B$ could be fairly assumed (within reasonable range) as

\begin{equation}\label{eq:lb}
    l_B = c_0 + c_1 k l
\end{equation}
where $c_0$ and $c_1$ are constant parameters specific to a model and hardware.

In the case of dynamic models, requests in a batch are padded to the largest one and
therefore, it can be viewed as if the whole batch's requests have the same length:

\begin{equation}
    l = \max_{r \in B} l_r
\end{equation}

Then the execution time of $B$ becomes a new random variable $L_B$.
With $L_r$ as the random variable of request $r$'s execution time, we have

\begin{equation}\label{eq:e-lb}
\begin{aligned}
    \E \lbrack L_B \rbrack &= c_0 + c_1 k \E \lbrack \max_{r \in B} l_r \rbrack  \\
     &= c_0 + c_1 k \int_0^\infty L_{r(k)} f_{L_{r(k)}}(l) \diff l
\end{aligned}
\end{equation}
where $L_{r(k)}$ is $L_r$'s max order statistics over $k$ samples, and $f_{L_{r(k)}}(l)$ is its PDF\@.
While it is possible to directly find out $f_{L_{r(k)}}(l)$, it is easier to go through the cumulative distribution function (CDF) of $L_{r(k)}$ first, denoted as $F_{L_{r(k)}}(l)$, which is related to the CDF of $L_r$ via a simple equation:

\begin{equation}
F_{L_{r(k)}}(l) = \lbrack F_{L_r}(l) \rbrack^k
\end{equation}

And we can obtain $F_{L_r}(l)$ from $L_r$'s histogram.
Note that while it is possible to directly use $F_{L_{r(k)}}(l)$ to calculate $\E \lbrack \max_{r \in B} l_r \rbrack$, the result would be far too inaccurate, as we only have a discrete histogram to start with.

\subsubsection{Requests in Batch Follow Different Distributions}%
\label{sec:algo-diff}

Taking one step further, if requests $r_i (i = 1, 2, \dots, k)$ in $B$ come from different distributions, the problem becomes finding the max order statistics $L_{r(k)}$ for $k$ random variables $L_{r_i}$ that are independent, but not necessarily identically distributed.

Let $F_i$ and $f_i$ be the CDF and PDF for $L_{r_i}$, respectively, and define $F^s, f^s$ as

\begin{equation}
    \begin{aligned}
        F^s &= \frac{1}{n_s} \sum_{i \in s} F_i \\
        f^s &= \frac{1}{n_s} \sum_{i \in s} f_i
    \end{aligned}
\end{equation}
where $s$ is a subset of requests ($s \subseteq B$) with $n_s \ge 1$ elements.
The PDF of the maximum (\ie{} $k$-th order statistics) of these $k$ variables $f_{(k)}$ is given by Özbey \etal~\cite{max-innid}:

\begin{equation}\label{eq:pdf-innid}
    f_{L_{r(k)}} =
        \sum^k_{\kappa = 1} (-1)^{k-\kappa} \frac{\kappa^k}{k!}
        \sum_{n_s = \kappa}
            k \lbrack F^s \rbrack^{k-1} f^s
\end{equation}

Here $\sum_{n_s = \kappa}$ means summation over all possible $s \subseteq B$ where $n_s = \kappa$.

With $f_{L_{r(k)}}$, we can describe the PDF of $L_B$ using \autoref{eq:lb}
\begin{equation}\label{eq:pdf-lb}
    f_{L_B}(l) = f_{L_{r(k)}}(\frac{l - c_0}{c_1k})
\end{equation}

Similarly, by plugging \autoref{eq:pdf-innid} into \autoref{eq:e-lb}, we can also find out $\E \lbrack L_B \rbrack$.
We now have complete ingredients to compute the priority score $p(t)$.

\begin{figure}
    \centering
    \subcaptionbox{
        Execution time of two types of requests.%
        \label{fig:toy-example-2dist}
    }[.45\columnwidth]{
        \includegraphics{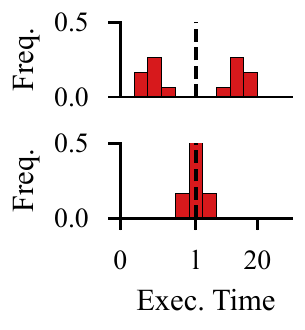}
    }\hfill
    \subcaptionbox{
        Execution time histogram for the batch.%
        \label{fig:toy-example-batch-dist}
    }[.45\columnwidth]{
        \includegraphics{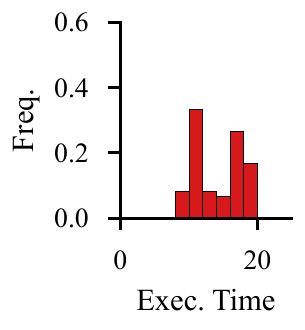}
    }
    \subcaptionbox{
        The priority score of $r_1, r_2, r_3$ entering the system one after another.%
        \label{fig:toy-example-score}
    }{
        \includegraphics{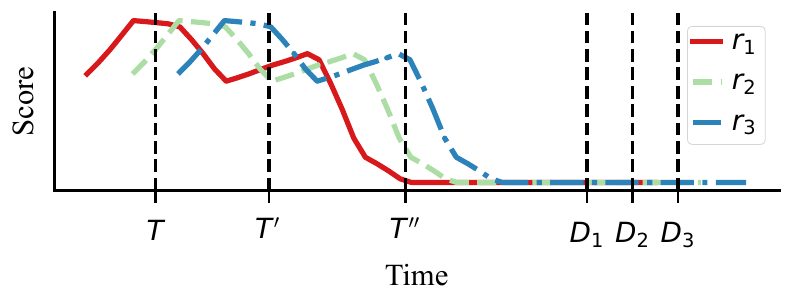}
    }
    \caption{Toy example of batch execution time estimation and priority score computation.}%
    \label{fig:toy-example}
\end{figure}

\paragraph{A Toy Example}%
\label{sec:example}
Let us use an example to illustrate how $p(t)$ changes over time $t$, to put the above discussed equations into perspective.

Consider two types of requests in total, whose execution time follow two different distributions, as shown in \autoref{fig:toy-example-2dist}.
While they all have the same mean execution time $l$, the first distribution has higher probability of finishing exactly at $l$, and the second one may either finish very early, or very late.
Further, assuming that the batch size in consideration is 2 and there is no overhead for batching (\ie{} $c_0 = 0, c_1k = 1$).

\autoref{fig:toy-example-batch-dist} shows the histogram of $L_B$ that is derived using \autoref{eq:pdf-lb}.
As expected, the execution time for the whole batch is skewed to the right, because it is not possible for the shorter execution time in the first distribution to ever become the whole batch's execution time.

\autoref{fig:toy-example-score} illustrates the priority score changing over time for three requests $r_1, r_2, r_3$ entering the system one after another.
At time $T$, $r_1$ and $r_2$ are the most urgent and should be executed.
As the deadline is approaching, $r_1$ and $r_3$ become the top ones at time $T'$.
Finally, $r_2$ and $r_3$ would have been selected as the score for $r_1$ drops to 0.

\subsection{Batch Formation}%
\label{sec:batch-formation}

The last step is forming the batch, which must address a key challenge: the circular dependency between priority calculation and batch formation.
$\E \lbrack L \rbrack$ is used in the priority calculation, so the list of possible distributions must be known.
However, the batch is determined \emph{after} priority is computed, and thus it is impossible to know what other requests are in the batch.
Assuming that the queue most likely contain requests from all types of applications the model is serving,
we therefore always use all execution time distributions associated with the model to compute $\E \lbrack L \rbrack$.
As this list of distributions is available ahead of time, and only depends on the batch size, this approach has the additional benefit that the relatively heavy computation can be moved away from the critical path.

\subsection{Efficient Computation}%
\label{sec:convex-hull}

As shown before, the priority score for a request varies depending on the remaining time before its deadline.
It thus has to be re-computed continuously.
Combined with the effort to re-sort all pending requests, the naive implementation is not scalable to large number of requests.

It is possible to convert the problem to dynamic convex hull querying, which can achieve $O(\log^2 n)$ complexity for each reschedule, where $n$ is the number of pending requests~\cite{icbs}.
The dynamic convex hull problem is defined by rewrite each request's priority \autoref{eq:pt} in the form of $p_i(t) = \alpha e^{bt} + \beta$, and consider each request to be on a 2D plane with the coordinate $(\alpha, \beta)$.
Then, the first point on the convex hull hit by affine lines of slope $-e^{bt}$ corresponds to the request with the highest score at time $t$.

This way, the problem is divided into one time-invariant part where the relative positions of requests only change a few times (when the relationship between $t, D, l^{(i)}_1, l^{(i)}_2$ changes, corresponding to the \texttt{Milestone} function in \autoref{alg:dyninfer-iter}), and one time-varying part -- querying the convex hull with a line.
Therefore, the priority queue can be implemented by maintaining a convex hull containing all pending requests.

However, while querying a static convex hull is trivial, our convex hull changes over time as requests come and go.
In \name{}, we use the convex hull algorithm proposed by Overmars and von Leeuwen~\cite{dynamic-convex-hull}, which supports dynamically adding/removing points on the convex hull in $O(\log^2 n)$ complexity and can be queried with a line in $O(1)$ time.

\paragraph{Overflow Handling of Exponential Values}

While the theory works out, there is still a non-trivial challenge that we faced when implementing the priority score in practice.

In the original \autoref{eq:pt}, the score only depends on $D - t$ which is the remaining time before the deadline, and is bounded assuming requests too far in the future should not enter the system.
However, the clever 2D plane mapping breaks the component into $e^{-bD}$ and $e^{bt}$ individually.
Because $D$ and $t$ can be large timestamps (usually represented as elapsed seconds/milliseconds since UNIX epoch), this leads to floating-point overflow when the system tries to compute and store these very large exponential values.

We compensate this by using relative timestamps for $D$ and $t$, and then choose $b$ wisely.
If the time resolution is in milliseconds, and $b = 10^{-4}$,
we can sustain about $1000~\text{s}$ of scheduling before overflows of 64-bit floating-point numbers and having to reset the relative timestamps' reference point and thus re-calculate everything.
Note that the exact value of $b$ does not matter because it does not change the relative ordering of requests as long as it is kept constant.

\section{Evaluation}%
\label{sec:eval}

\begin{table*}[!t]
    \centering
    \caption{List of workloads.}%
    \label{tbl:workloads}
    \begin{tabular}{rllrr}
        \toprule
        Task  & Model & Dataset & Mean Exec. (ms) & P99 Exec. (ms) \\
        \midrule
        Image classification	& RDI-Nets~\cite{rdi-nets}	& CIFAR~\cite{cifar}	&  683.15	& 2667.54 \\
        Image classification	& SkipNet~\cite{skipnet}	& ImageNet~\cite{imagenet}	&  3.24	& 5.56 \\
        \midrule
        Chatbot			& Blenderbot~\cite{blenderbot}	& convAI~\cite{convai}	&  200.39	& 242.27 \\
        Chatbot			& Blenderbot	& Cornell~\cite{cornell}	&  203.22	& 247.04 \\
        Chatbot			& GPT~\cite{gpt}		& convAI	&  
79.47 & 143.40 \\
        Chatbot			& GPT		& Cornell	&  94.84 & 161.69 \\
        Summarization	& BART~\cite{bart}		& CNN~\cite{cnn}	& 774.66 & 1101.99 \\
        Summarization	& T5~\cite{t5}		& CNN	& 552.91 & 797.28 \\
        Translation		& FSMT~\cite{fsmt}		& WMT~\cite{wmt}	&  189.30 & 319.31 \\
        Translation		& mBART~\cite{mbart}		& WMT	&  432.38 & 729.87 \\
        \bottomrule
    \end{tabular}
\end{table*}

\begin{figure*}[!t]
    \centering
    \includegraphics{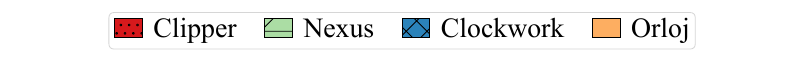}

    \subcaptionbox{
        Image classification/RDI-Nets/CIFAR%
        \label{fig:finish-rate-rdinet_cifar}%
    }{
        \includegraphics{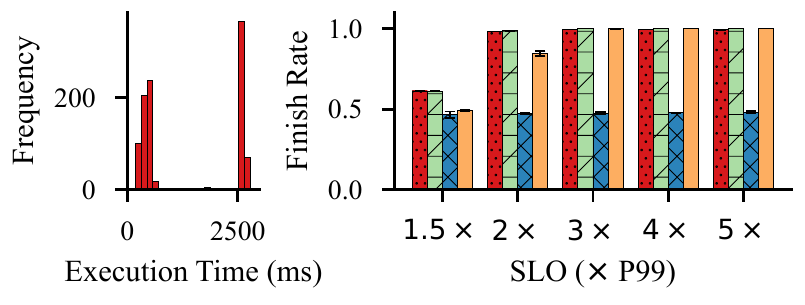}
    }
    \subcaptionbox{
        Summarization/BART/CNN%
        \label{fig:finish-rate-bart_cnn}%
    }{
        \includegraphics{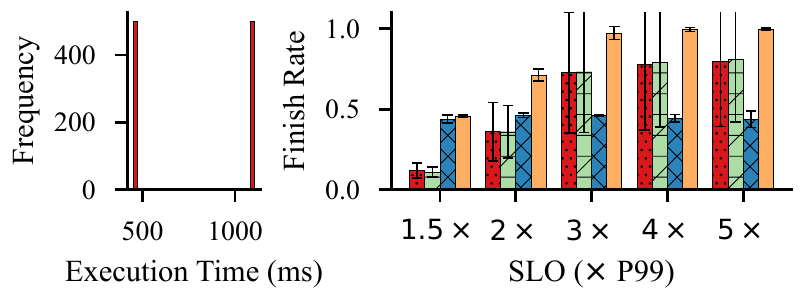}
    }
    \subcaptionbox{
        Image classification/SkipNet/ImageNet%
        \label{fig:finish-rate-skipnet_imagenet}%
    }{
        \includegraphics{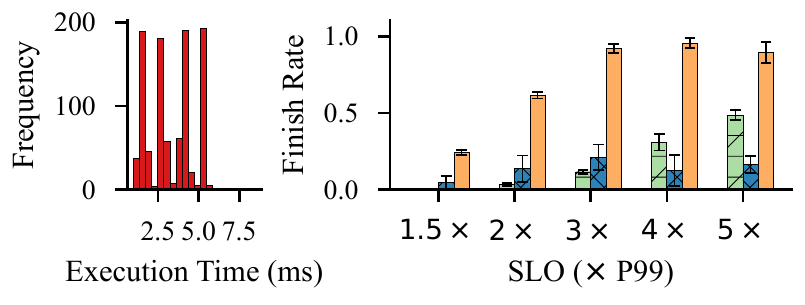}
    }
    \subcaptionbox{
        Chatbot/GPT/Cornell%
        \label{fig:finish-rate-gpt_cornell}%
    }{
        \includegraphics{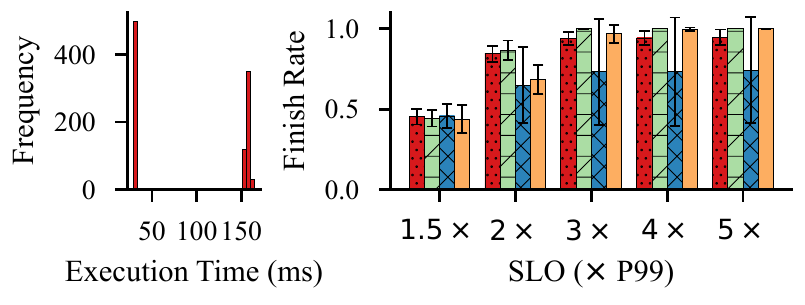}
    }
    \caption{\name{} performance on real world tasks. Error bars represent standard deviation across 5 runs.}%
    \label{fig:real-tasks}
\end{figure*}

We evaluate \name{} against three existing serving systems (Clipper~\cite{clipper}, Nexus~\cite{nexus}, and Clockwork~\cite{clockwork}).
Our primary findings are as follows:
\begin{itemize}
    \item Compared to the state-of-the-art, \name{} can improve the finish rate when serving dynamic DNNs by 51--80\% under extremely tight SLO constrains, or over 100\% under more relaxed SLO settings (\autoref{sec:eval-dyn}).
    At the same time, \name{} keeps comparable performance when serving traditional static models (\autoref{sec:eval-static}).

    \item \name{} can sustain thousands of pending requests in its priority queue with less than $0.5~\text{ms}$ per-request insertion time (\autoref{sec:eval-sensitivity}).

    \item Our choice of $b$, the anticipated delay distribution parameter (\autoref{sec:algo-prelim}) in the priority score is safe, as \name{} is not sensitive to the value of $b$ (\autoref{sec:eval-lambdas}).

    \item \name{} has minimal overheads and can manage requests with execution time varying in ranges as low as $2~\text{ms}$--$20~\text{ms}$ (\autoref{sec:eval-overheads}).
\end{itemize}

\subsection{\name{} Implementation}
We implemented \name{} on top of Clockwork~\cite{clockwork}, a state-of-the-art serving system with fewer than 4000 lines of C++ code.
One-fourth of the new code is for implementing the dynamic convex hull data structure, as there is no established library available for solving the dynamic convex hull problem.
Specifically, we implemented the inner \emph{concatenate queue} as a 2--3 tree extending from the left-leaning-red-black-tree~\cite{concatqueue-1, concatqueue-2}.

\subsection{Experimental Methodology}

\paragraph{Testbed}
We built our testbed on Chameleon Cloud~\cite{chameleon}.
The host has 2 Intel Xeon Gold 6230 CPUs with NVIDIA V100 GPU\@.
In order to have stable results, we fix the GPU clock speed to its maximum $1380~\text{MHz}$ and memory clock speed to $877~\text{MHz}$.

We use Ubuntu 20.04 as the base OS environment with the latest NVIDIA GPU driver.
We use CUDA versions matching the published original source code, which means CUDA 11.1.1 with CuDNN 8 for Clockwork and \name{}, CUDA 10.0 with CuDNN 7 for Nexus.
For Clipper, its latest commit \texttt{9f25e3f} is used.

During experiments, each evaluated system's server is deployed on the host.
Clockwork and \name{} additionally have their serving threads set to high-priority and pinned to physical cores as per Clockwork's host setup instructions.
In addition, the model is modified to allow us to explicitly control its execution time via input for the purpose of evaluation.

An open loop (no wait for requests completion before issuing the next one) client is used to drive all experiments on the same host to minimize the impact of networking.

\paragraph{Input Trace}

Similar to workload traces for static models, the request incoming rate trace determines how fast requests coming in and needs to match the system's load.
Same as Clockwork's evaluation, we adapt the Azure trace~\cite{azure}, which is published by Microsoft for lambda functions.
The trace was scaled down such that the incoming rate matches the system load.
The incoming rate trace is kept the same across all experiments.

\paragraph{Request Execution Time Distribution}
Unlike a single number for one model/dataset combination in evaluations in static serving systems, we need a full distribution.
We group the model's associated dataset into short-running and relatively long-running requests (or more groups in case of higher modality), then randomly choose from them to get a mixture of both.
To get a fair comparison, the generation is done once among different runs, we then
record the arrival time and the input, which will be replayed for subsequent runs.

\paragraph{Real World Dataset}

We evaluate \name{} on a set of real world learning tasks covering both CV ones like image classification, and NLP ones including chatbot, summarization and translation (\autoref{tbl:workloads}).
For each workload, the input execution time distribution is determined according to the above-mentioned method, and the P99 of real execution time used to determine SLO is reported in the table.

\paragraph{Metrics}
We focus on the \emph{finish rate}, which is defined as the ratio of the number of requests finished in time to the total number of requests.
We assume that the SLO is set to a reasonable value manually given historical data, similar to serving static models.
Using the P99 tail of all input requests' real execution time as a measurement, we vary the SLO to be multiples of P99 for most our experiments.

\subsection{Improvements for Dynamic Workloads}%
\label{sec:eval-dyn}

\paragraph{Real World Dataset}

We report representative cases in \autoref{fig:real-tasks}, while the complete results can be found in \autoref{sec:appendix}.
In most workloads, existing systems can barely make it due to the mixture of long and short requests which rarely match the mean execution time those systems use for scheduling.
For RDI-Nets/CIFAR, BART/CNN and GPT/Cornell, \name{} can reach near 100\% finish rate with sufficient SLO settings.
When requests become extremely short (\eg, \autoref{fig:finish-rate-skipnet_imagenet}), none of the system can finish many requests due to the too tight latency requirement.
However, \name{} still manages to finish more requests than others.

\paragraph{Different Distributions}
We then evaluate \name{}'s performance under more diverse distributions using the same BART model and with a synthesized dataset to control execution times.

\begin{figure}[!t]
    \centering
    \captionbox{
        Finish rate under different modality distributions.\label{fig:diff-dist-modal}
    }[\columnwidth]{
        \includegraphics{finish-rate-legend}
        \subcaptionbox{
            Simple normal distribution%
            \label{fig:finish-rate-one_modal}%
        }{
            \includegraphics{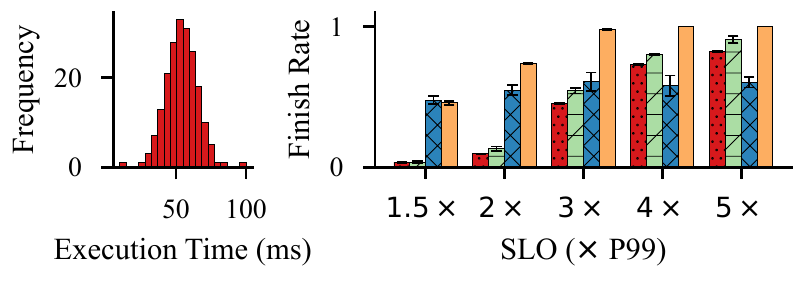}
        }
        \subcaptionbox{
            Bimodal distribution%
            \label{fig:finish-rate-two_modal}%
        }{
            \includegraphics{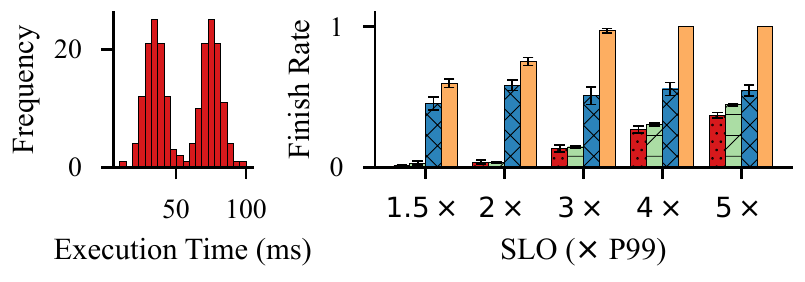}
        }
        \subcaptionbox{
            Three-modal distribution%
            \label{fig:finish-rate-three_modal}%
        }{
            \includegraphics{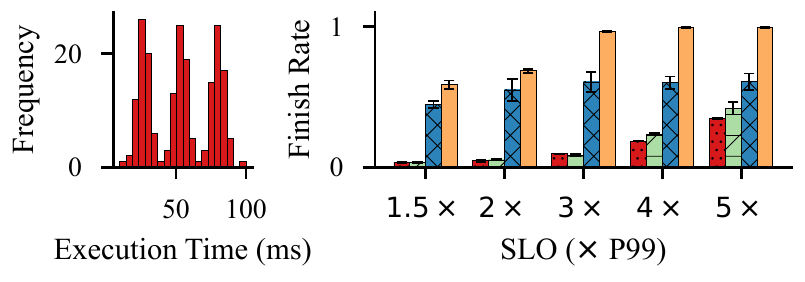}
        }
    }

    \vspace*{1em}

    \captionbox{
        \name{}'s performance under inequal-peak distributions.\label{fig:two-peaks}
    }[\columnwidth]{
        \includegraphics{finish-rate-legend}
        \subcaptionbox{
            Bimodal distribution with inequal peaks (more short requests)%
            \label{fig:finish-rate-left_2_right_0.5}%
        }{
            \includegraphics{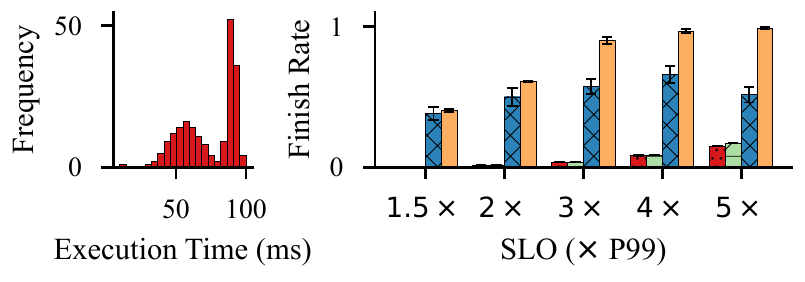}
        }
        \subcaptionbox{
            Bimodal distribution with inequal peaks (more long requests)%
            \label{fig:finish-rate-left_0.5_right_2}%
        }{
            \includegraphics{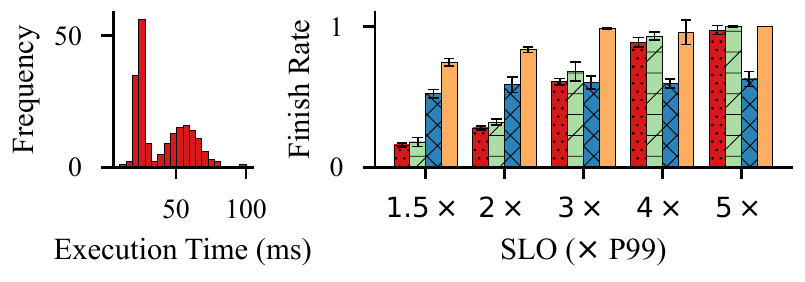}
        }
    }
\end{figure}

In \autoref{fig:diff-dist-modal}, we increase the number of modalities of the distribution to simulate the effect of multiple applications.
With the number of modalities increases, the variation in execution time increases.
\name{} keeps relatively good finish rate and see performance gain as high as $2\times$.
The result is consistent with even higher modalities, and we report additional results for up to 8 modals in \autoref{sec:appendix}.

The distributions in \autoref{fig:two-peaks} are the same except for mirrored inequal peak locations.
However, Clipper and Nexus suffer more in \autoref{fig:finish-rate-left_2_right_0.5} because exceptional longer than expected requests are definitely timed out while exceptional shorter than expected requests can still meet the deadline.

\begin{figure}[!t]
    \centering
    \includegraphics{finish-rate-legend}
    \subcaptionbox{
        Bimodal distribution with $\sigma = 0.5$%
        \label{fig:finish-rate-std_0.5}%
    }{
        \includegraphics{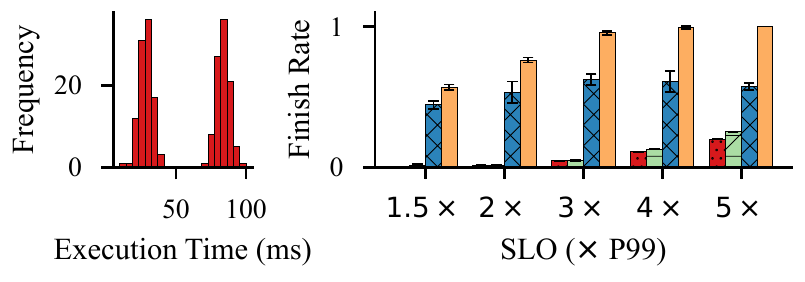}
    }
    \subcaptionbox{
        Bimodal distribution with $\sigma = 2$%
        \label{fig:finish-rate-std_2}%
    }{
        \includegraphics{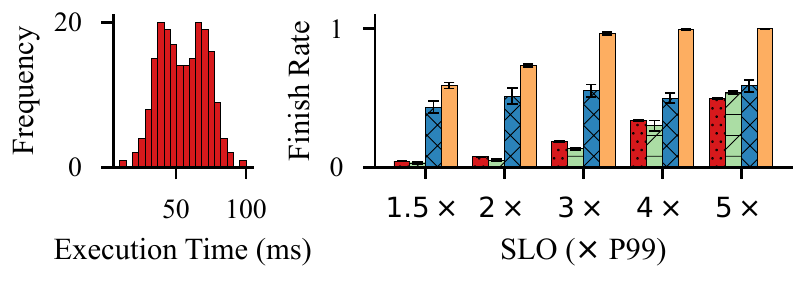}
    }
    \caption{Finish rate under different distribution parameters.}%
    \label{fig:diff-dist-std}
\end{figure}

\autoref{fig:diff-dist-std} extends on \autoref{fig:finish-rate-two_modal}, whose input request execution time distribution is generated with $\sigma = 1$, and explores the effect of smaller ($\sigma = 0.5$) and larger ($\sigma = 1$) values.
Larger $\sigma$ means the peaks are less distinguishable and the degrees of longer requests blocking shorter ones is less severe.
\name{}'s performance remains stable while others see slightly higher finish rate when the separation between requests become less extreme and vice versa.

Clockwork's performance is not affected by changes in distributions. As long as the execution time is not constant, it suffers from the same fail-every-other-batch pattern as we discussed in \autoref{sec:motivation}.

\subsection{Improvements for Traditional Workloads}%
\label{sec:eval-static}
We next verify \name{}'s performance under traditional workloads using static models where there is no variance in request execution time.

\begin{figure}
    \centering
    \includegraphics{finish-rate-legend}
    \subcaptionbox{
        Inception model on ImageNet%
        \label{fig:finish-rate-inception_imagenet}%
    }{
        \includegraphics{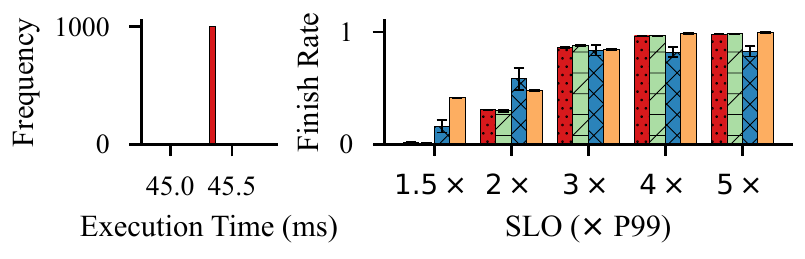}
    }
    \subcaptionbox{
        ResNet model on ImageNet%
        \label{fig:finish-rate-resnet_imagenet}%
    }{
        \includegraphics{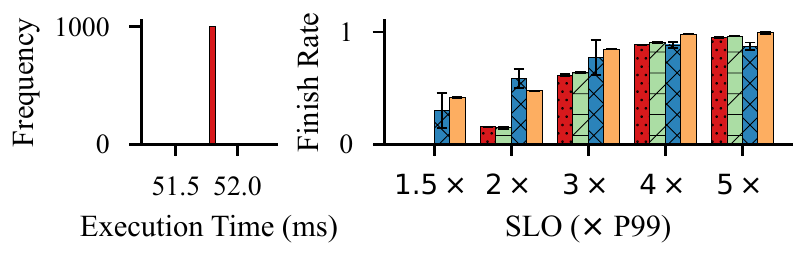}
    }
    \caption{\name{} keeps comparable performance under workloads where there is no variance in request execution time.}%
    \label{fig:finish-rate-cv}
\end{figure}

Using the ImageNet~\cite{imagenet} dataset, we measure the finish rate for four systems when serving the ResNet~\cite{resnet} model and Inception V3~\cite{inception} model (\autoref{fig:finish-rate-cv}).
\name{} sees significant improvement over Nexus and Clipper under tight SLOs ($1.5\times$ and $2\times$), thanks to its plan-ahead scheduling.
When compared to Clockwork, upon which \name{} is built, due to differences in request handling mechanisms, \name{} performs slightly better when SLO is higher while Clockwork has higher finish rate under tight SLOs, albeit with higher variances.


\subsection{Efficiency of the Priority Queue}%
\begin{figure}
    \centering
    \includegraphics{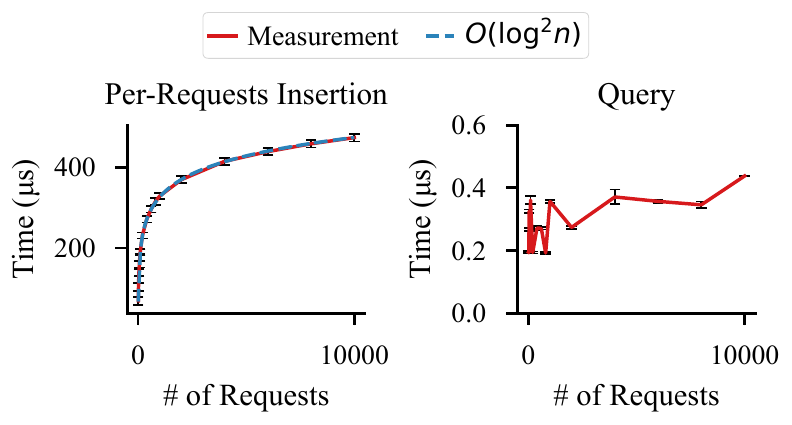}
    \caption{The insertion time of \name{} priority queue under different numbers of requests in queue.}%
    \label{fig:eval-queue-complexity}
\end{figure}

\label{sec:eval-sensitivity}
To study the efficiency of our priority queue implementation, we evaluate two of the most common operation of the priority queue in isolation: insertion and query.
To measure the time it takes to insert a request into the queue, we run micro-benchmarks that fill the queue to certain number of requests, and compute the average insertion time per-request.
For query, we first fill the queue and measure the time it takes to query the queue against a line of random slope -- the equivalent of finding the request with the highest priority, as discussed in \autoref{sec:convex-hull}.
We vary the number of requests in queue from 10 to 10000, and each data point is averaged over 100 samples.
As reported in \autoref{fig:eval-queue-complexity}, the insertion operation takes longer as the queue becomes larger, and the overall complexity trend fits the theory $O(\log^2 n)$ line pretty well.
Query time sees large variation when the number of requests is small, but stabilizes and remains constant as the queue size increases.
Overall, we can see that thanks to the efficient implementation, thousands of requests can be handled in negligible time, and thus \name{} is able to schedule large number of pending requests.

\subsection{Sensitivity to the Anticipated Delay Distribution}%
\label{sec:eval-lambdas}

In the priority score calculation in \autoref{sec:algo-prelim}, we introduce a parameter $b$ when describing the distribution of the anticipated delay.
And we discussed that in order to avoid floating-point overflow during calculation, we choose $b = 10^{-4}$ in \autoref{eq:pt}.
To verify that our choice of $b$ is reasonable and \name{}'s scheduling is not sensitive to the value of $b$, we do a parameter sweep with $b = 10^{-6}, 10^{-5}, \dots, 10^{-1}$, and measure the performance of \name{} using the three-modal distribution as shown in \autoref{fig:finish-rate-three_modal}.

\begin{figure}
    \centering
    \includegraphics{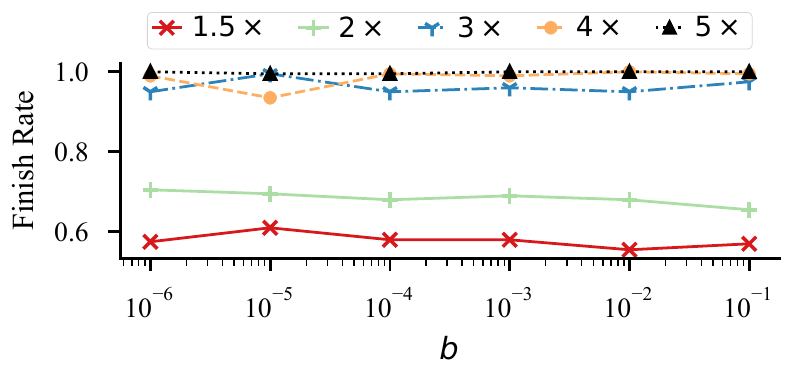}
    \caption{Finish rate as we vary $b$. Note the x-axis is in log scale.}%
    \label{fig:eval-lambdas}
\end{figure}

Each line in \autoref{fig:eval-lambdas} represents the trend of finish rate under a given SLO setting (as a multiple of P99 execution time).
And it can be seen that under all SLO settings, \name{} indeed keeps stable finish rate regardless of the choice of $b$.
It is therefore safe to choose $b$ to account for floating-point overflow, as we do in \name{}'s implementation.

\subsection{Overheads}%
\label{sec:eval-overheads}

To understand \name{}'s scheduling overhead, we evaluate the lower limit on SLOs that \name{} can achieve by measuring the finish rate while varying incoming requests' minimum execution time (time to execute one request alone).
In this experiment, we use the same workload with three-modal distribution as shown in \autoref{fig:finish-rate-three_modal},
and scale the whole execution time distribution down until \name{}'s finish rate drops significantly.

\begin{figure}
    \centering
    \includegraphics{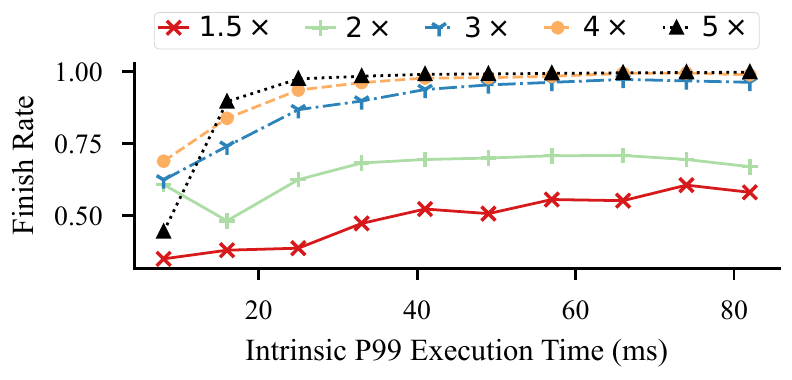}
    \caption{Finish rate as we vary incoming requests' minimum execution time.}%
    \label{fig:eval-overheads}
\end{figure}

\autoref{fig:eval-overheads} reports the trends of finish rates under different SLO settings.
Similar to other experiments, we set SLOs according to the P99 of minimum execution time and report results under different SLO to P99 ratios from $1.5\times$ to $5\times$.
\name{} keeps consistent and stable finish rate. 
Its performance only starts to degrade when the minimum P99 execution time is approaching $20~\text{ms}$. 
At that time, due to variance in request execution time, the mean approaches $10~\text{ms}$ and requests can go as low as less than $2~\text{ms}$.

\section{Related Work}

\paragraph{Model Serving}

We directly compare \name{} to Clockwork~\cite{clockwork}, Nexus~\cite{nexus} and Clipper~\cite{clipper}.
While Clockwork's assumption about static DNNs no longer holds for the dynamic DNNs, its idea of proactively planning ahead and consolidating choices across layers to reduce disturbance still applies in \name{}.
In addition, existing systems propose several orthogonal concepts that can be seen as complementary to \name{}.
Clipper's idea of model selection and INFaaS's~\cite{infaas} model variant concept could be applied in front of \name{}.
Nexus' model prefix-sharing and InferLine's~\cite{inferline} inference pipelines are also compatible with \name{}'s idea of tracking request execution time as random variables.

In cloud or serverless platforms there are projects focusing on serving models at scale~\cite{serving1,serving2,serving3}.
TensorFlow Serving~\cite{tfserving} provides one of the first production environments for models trained using the TensorFlow framework.
SageMaker~\cite{sagemaker}, Vertex AI~\cite{vertex-ai} and Azure ML~\cite{azure-ml} are public cloud DNN serving systems that offer developers inference services that auto-scale based on load.

\paragraph{Cost-Aware Scheduling}
One family of the well-studied scheduling algorithms are cost-aware or utility-based algorithms.
The decisions in these algorithms are made to optimize certain costs, which could be defined in many ways: they could be fixed or time-varying values~\cite{sched1,sched2}, costs of rolling back transactions~\cite{sched3}, or derived from SLOs~\cite{slo1,slo2,slo3}.
It is however common in these algorithms to assume the exact execution time to be available during scheduling, which is not the case in serving dynamic DNNs.

\paragraph{Unknown-Sized Job Scheduling in Cluster}
Scheduling jobs with unknown duration has been studied in cluster computing.
3Sigma~\cite{3sigma} uses the job length distribution in the scheduling and enumerates all possible choices to find the optimal scheduling decision.
Age-based mechanisms~\cite{tiresias:nsdi19, soap} gradually update jobs' priorities based on sustained service time.
There are also various techniques used to mitigate mis-prediction in case the job's length exceeds expectation~\cite{optimus}.

However, inference request serving differs from the above in its timescale and properties.
Inference requests usually complete in less than a second, whereas cluster jobs can last hours or even days.
So our scheduler has to make decision very quickly, unlike cluster schedulers that can use extensive searching before settling on a schedule.
Furthermore, while cluster jobs are usually preemptable, inference requests are not, which rules out age-based algorithms.

\section{Conclusion}

Dynamic DNNs adapt their structures or parameters to the input, and thus experiencing rapid development thanks to notable advantages in terms of accuracy, computational efficiency, adaptiveness, \etc
This challenges existing DNN serving solutions that assume data-independence of incoming requests, and they suffer from poor performance due to the large variance in request execution times.
We propose \name{}, a dynamic DNN serving system, to meet these challenges.
\name{} captures the variance in dynamic DNNs by modeling request execution time as random variables, and then efficiently batches and schedules them without knowing a request's precise execution time.
We demonstrated that \name{} significantly outperforms states-of-the-art serving solutions for high variance dynamic DNN workloads while maintaining nearly identical performance for static workloads.

While we take a first step in this paper, we hope that \name will inspire further research not only on dynamic DNN inference serving systems but other aspects of dynamic DNN lifecycle as well.

\clearpage
\printbibliography{}

\clearpage
\appendix
\section{Additional Experiment Results}%
\label{sec:appendix}

We report more results in \autoref{tbl:res-std}, \autoref{tbl:res-modality}, \autoref{tbl:res-cv}, and \autoref{tbl:res-real}.

\begin{table}[!b]
    \centering
    \caption{Evaluation results for cases where request execution time distribution is bimodal. Case ID shows the standard deviation of each modal's normal distribution.}%
    \label{tbl:res-std}
    \begin{small}
    \setlength\tabcolsep{1.5pt} 
\begin{tabular}{cccccc}
\toprule
Case ID & SLO & \multicolumn{4}{c}{Finish Rate}          \\
        & ($\times$ P99)  &       Clipper & Nexus & Clockwork & \name{} \\
\midrule
        \texttt{std-0.5} & 1.5 & 0.01 & 0.01 & 0.44 & \textbf{0.57} \\
        \texttt{std-0.5} & 2 & 0.01 & 0.02 & 0.53 & \textbf{0.76} \\
        \texttt{std-0.5} & 3 & 0.04 & 0.05 & 0.62 & \textbf{0.96} \\
        \texttt{std-0.5} & 4 & 0.11 & 0.13 & 0.61 & \textbf{0.99} \\
        \texttt{std-0.5} & 5 & 0.20 & 0.25 & 0.57 & \textbf{1.00} \\
        \texttt{std-1} & 1.5 & 0.02 & 0.02 & 0.46 & \textbf{0.60} \\
        \texttt{std-1} & 2 & 0.03 & 0.03 & 0.53 & \textbf{0.76} \\
        \texttt{std-1} & 3 & 0.10 & 0.10 & 0.55 & \textbf{0.97} \\
        \texttt{std-1} & 4 & 0.21 & 0.19 & 0.53 & \textbf{0.99} \\
        \texttt{std-1} & 5 & 0.33 & 0.34 & 0.56 & \textbf{1.00} \\
        \texttt{std-2} & 1.5 & 0.04 & 0.03 & 0.43 & \textbf{0.59} \\
        \texttt{std-2} & 2 & 0.07 & 0.05 & 0.51 & \textbf{0.73} \\
        \texttt{std-2} & 3 & 0.19 & 0.13 & 0.55 & \textbf{0.97} \\
        \texttt{std-2} & 4 & 0.34 & 0.30 & 0.50 & \textbf{1.00} \\
        \texttt{std-2} & 5 & 0.49 & 0.54 & 0.59 & \textbf{1.00} \\
        \texttt{std-2/0.5} & 1.5 & 0.16 & 0.18 & 0.52 & \textbf{0.75} \\
        \texttt{std-2/0.5} & 2 & 0.28 & 0.32 & 0.59 & \textbf{0.83} \\
        \texttt{std-2/0.5} & 3 & 0.61 & 0.68 & 0.60 & \textbf{0.98} \\
        \texttt{std-2/0.5} & 4 & 0.89 & 0.93 & 0.59 & \textbf{0.96} \\
        \texttt{std-2/0.5} & 5 & 0.97 & 1.00 & 0.63 & \textbf{1.00} \\
        \texttt{std-0.5/2} & 1.5 & 0.01 & 0.01 & 0.38 & \textbf{0.40} \\
        \texttt{std-0.5/2} & 2 & 0.01 & 0.02 & 0.50 & \textbf{0.61} \\
        \texttt{std-0.5/2} & 3 & 0.04 & 0.04 & 0.57 & \textbf{0.90} \\
        \texttt{std-0.5/2} & 4 & 0.08 & 0.09 & 0.66 & \textbf{0.97} \\
        \texttt{std-0.5/2} & 5 & 0.15 & 0.17 & 0.52 & \textbf{0.99} \\
\bottomrule
\end{tabular}

    \end{small}
\end{table}

\begin{table}
    \centering
    \caption{Evaluation results for cases where we vary the modality of request execution time distribution.}%
    \label{tbl:res-modality}
    \begin{small}
    \setlength\tabcolsep{1.5pt} 
\begin{tabular}{cccccc}
\toprule
Case ID & SLO & \multicolumn{4}{c}{Finish Rate}          \\
        & ($\times$ P99)  &       Clipper & Nexus & Clockwork & \name{} \\
\midrule
        \texttt{one-modal} & 1.5 & 0.03 & 0.04 & \textbf{0.48} & 0.46 \\
        \texttt{one-modal} & 2 & 0.10 & 0.13 & 0.55 & \textbf{0.74} \\
        \texttt{one-modal} & 3 & 0.45 & 0.54 & 0.61 & \textbf{0.98} \\
        \texttt{one-modal} & 4 & 0.73 & 0.80 & 0.58 & \textbf{1.00} \\
        \texttt{one-modal} & 5 & 0.82 & 0.91 & 0.60 & \textbf{1.00} \\
        \texttt{two-modal} & 1.5 & 0.01 & 0.03 & 0.45 & \textbf{0.60} \\
        \texttt{two-modal} & 2 & 0.04 & 0.03 & 0.58 & \textbf{0.75} \\
        \texttt{two-modal} & 3 & 0.13 & 0.14 & 0.51 & \textbf{0.97} \\
        \texttt{two-modal} & 4 & 0.27 & 0.30 & 0.56 & \textbf{1.00} \\
        \texttt{two-modal} & 5 & 0.37 & 0.44 & 0.55 & \textbf{1.00} \\
        \texttt{three-modal} & 1.5 & 0.03 & 0.04 & 0.45 & \textbf{0.59} \\
        \texttt{three-modal} & 2 & 0.05 & 0.05 & 0.55 & \textbf{0.68} \\
        \texttt{three-modal} & 3 & 0.10 & 0.09 & 0.61 & \textbf{0.97} \\
        \texttt{three-modal} & 4 & 0.18 & 0.23 & 0.60 & \textbf{0.99} \\
        \texttt{three-modal} & 5 & 0.35 & 0.42 & 0.61 & \textbf{0.99} \\
        \texttt{four-modal} & 1.5 & 0.03 & 0.05 & 0.46 & \textbf{0.59} \\
        \texttt{four-modal} & 2 & 0.05 & 0.07 & 0.56 & \textbf{0.73} \\
        \texttt{four-modal} & 3 & 0.08 & 0.12 & 0.59 & \textbf{0.94} \\
        \texttt{four-modal} & 4 & 0.10 & 0.16 & 0.60 & \textbf{0.98} \\
        \texttt{four-modal} & 5 & 0.14 & 0.19 & 0.60 & \textbf{1.00} \\
        \texttt{five-modal} & 1.5 & 0.02 & 0.03 & 0.47 & \textbf{0.60} \\
        \texttt{five-modal} & 2 & 0.02 & 0.03 & 0.55 & \textbf{0.73} \\
        \texttt{five-modal} & 3 & 0.03 & 0.07 & 0.59 & \textbf{0.94} \\
        \texttt{five-modal} & 4 & 0.04 & 0.08 & 0.57 & \textbf{0.97} \\
        \texttt{five-modal} & 5 & 0.07 & 0.08 & 0.56 & \textbf{0.99} \\
        \texttt{six-modal} & 1.5 & 0.04 & 0.04 & 0.46 & \textbf{0.58} \\
        \texttt{six-modal} & 2 & 0.03 & 0.05 & 0.53 & \textbf{0.71} \\
        \texttt{six-modal} & 3 & 0.06 & 0.08 & 0.53 & \textbf{0.92} \\
        \texttt{six-modal} & 4 & 0.07 & 0.09 & 0.51 & \textbf{0.97} \\
        \texttt{six-modal} & 5 & 0.08 & 0.11 & 0.54 & \textbf{0.99} \\
        \texttt{seven-modal} & 1.5 & 0.03 & 0.05 & 0.43 & \textbf{0.59} \\
        \texttt{seven-modal} & 2 & 0.04 & 0.05 & 0.51 & \textbf{0.73} \\
        \texttt{seven-modal} & 3 & 0.08 & 0.09 & 0.54 & \textbf{0.93} \\
        \texttt{seven-modal} & 4 & 0.10 & 0.12 & 0.52 & \textbf{0.98} \\
        \texttt{seven-modal} & 5 & 0.12 & 0.15 & 0.54 & \textbf{0.99} \\
        \texttt{eight-modal} & 1.5 & 0.03 & 0.04 & 0.33 & \textbf{0.60} \\
        \texttt{eight-modal} & 2 & 0.05 & 0.06 & 0.49 & \textbf{0.74} \\
        \texttt{eight-modal} & 3 & 0.08 & 0.09 & 0.43 & \textbf{0.93} \\
        \texttt{eight-modal} & 4 & 0.09 & 0.13 & 0.52 & \textbf{0.97} \\
        \texttt{eight-modal} & 5 & 0.11 & 0.14 & 0.50 & \textbf{0.99} \\
\bottomrule
\end{tabular}

    \end{small}
\end{table}

\begin{table}
    \centering
    \caption{Evaluation results for static models.}%
    \label{tbl:res-cv}
    \begin{small}
    \setlength\tabcolsep{1.5pt} 
\begin{tabular}{cccccc}
\toprule
Case ID & SLO & \multicolumn{4}{c}{Finish Rate}          \\
        & ($\times$ P99)  &       Clipper & Nexus & Clockwork & \name{} \\
\midrule
        \texttt{inception-imagenet} & 1.5 & 0.01 & 0.01 & 0.16 & \textbf{0.41} \\
        \texttt{inception-imagenet} & 2 & 0.31 & 0.30 & \textbf{0.58} & 0.48 \\
        \texttt{inception-imagenet} & 3 & 0.86 & \textbf{0.88} & 0.84 & 0.84 \\
        \texttt{inception-imagenet} & 4 & 0.96 & 0.97 & 0.82 & \textbf{0.99} \\
        \texttt{inception-imagenet} & 5 & 0.98 & 0.98 & 0.83 & \textbf{0.99} \\
        \texttt{resnet-imagenet} & 1.5 & 0.01 & 0.00 & 0.30 & \textbf{0.42} \\
        \texttt{resnet-imagenet} & 2 & 0.15 & 0.15 & \textbf{0.59} & 0.48 \\
        \texttt{resnet-imagenet} & 3 & 0.62 & 0.64 & 0.77 & \textbf{0.85} \\
        \texttt{resnet-imagenet} & 4 & 0.89 & 0.91 & 0.88 & \textbf{0.98} \\
        \texttt{resnet-imagenet} & 5 & 0.95 & 0.96 & 0.87 & \textbf{0.99} \\
\bottomrule
\end{tabular}

    \end{small}
\end{table}

\begin{table}
    \centering
    \caption{Evaluation results for real tasks. The first part of the case ID is the model name and second part the dataset name.}%
    \label{tbl:res-real}
    \begin{small}
    \setlength\tabcolsep{1.5pt} 
\begin{tabular}{cccccc}
\toprule
Case ID & SLO & \multicolumn{4}{c}{Finish Rate}          \\
        & ($\times$ P99)  &       Clipper & Nexus & Clockwork & \name{} \\
\midrule
        \texttt{blenderbot-convAI} & 1.5 & 0.05 & 0.06 & 0.43 & \textbf{0.45} \\
        \texttt{blenderbot-convAI} & 2 & 0.18 & 0.21 & 0.61 & \textbf{0.65} \\
        \texttt{blenderbot-convAI} & 3 & 0.56 & 0.63 & 0.64 & \textbf{0.74} \\
        \texttt{blenderbot-convAI} & 4 & 0.71 & 0.81 & 0.65 & \textbf{0.81} \\
        \texttt{blenderbot-convAI} & 5 & 0.75 & \textbf{0.87} & 0.65 & 0.81 \\
        \texttt{blenderbot-cornell} & 1.5 & 0.06 & 0.06 & 0.43 & \textbf{0.44} \\
        \texttt{blenderbot-cornell} & 2 & 0.20 & 0.20 & 0.60 & \textbf{0.68} \\
        \texttt{blenderbot-cornell} & 3 & 0.54 & 0.57 & 0.62 & \textbf{0.75} \\
        \texttt{blenderbot-cornell} & 4 & 0.71 & 0.74 & 0.64 & \textbf{0.81} \\
        \texttt{blenderbot-cornell} & 5 & 0.74 & 0.77 & 0.64 & \textbf{0.82} \\
        \texttt{gpt-convAI} & 1.5 & \textbf{0.39} & 0.38 & 0.39 & 0.36 \\
        \texttt{gpt-convAI} & 2 & 0.79 & \textbf{0.83} & 0.57 & 0.64 \\
        \texttt{gpt-convAI} & 3 & 0.91 & \textbf{0.99} & 0.61 & 0.86 \\
        \texttt{gpt-convAI} & 4 & 0.92 & \textbf{1.00} & 0.63 & 0.97 \\
        \texttt{gpt-convAI} & 5 & 0.92 & \textbf{1.00} & 0.61 & 0.98 \\
        \texttt{gpt-cornell} & 1.5 & 0.45 & 0.44 & \textbf{0.46} & 0.44 \\
        \texttt{gpt-cornell} & 2 & 0.84 & \textbf{0.86} & 0.65 & 0.68 \\
        \texttt{gpt-cornell} & 3 & 0.94 & \textbf{1.00} & 0.73 & 0.97 \\
        \texttt{gpt-cornell} & 4 & 0.94 & \textbf{1.00} & 0.73 & 0.99 \\
        \texttt{gpt-cornell} & 5 & 0.95 & 1.00 & 0.74 & \textbf{1.00} \\
        \texttt{bart-cnn} & 1.5 & 0.12 & 0.11 & 0.44 & \textbf{0.46} \\
        \texttt{bart-cnn} & 2 & 0.36 & 0.36 & 0.46 & \textbf{0.71} \\
        \texttt{bart-cnn} & 3 & 0.73 & 0.73 & 0.46 & \textbf{0.97} \\
        \texttt{bart-cnn} & 4 & 0.78 & 0.79 & 0.44 & \textbf{0.99} \\
        \texttt{bart-cnn} & 5 & 0.80 & 0.81 & 0.44 & \textbf{1.00} \\
        \texttt{t5-cnn} & 1.5 & 0.48 & 0.46 & 0.47 & \textbf{0.49} \\
        \texttt{t5-cnn} & 2 & \textbf{0.86} & 0.84 & 0.50 & 0.74 \\
        \texttt{t5-cnn} & 3 & 0.99 & 1.00 & 0.50 & \textbf{1.00} \\
        \texttt{t5-cnn} & 4 & 0.99 & 1.00 & 0.52 & \textbf{1.00} \\
        \texttt{t5-cnn} & 5 & 0.99 & 1.00 & 0.51 & \textbf{1.00} \\
        \texttt{fsmt-wmt} & 1.5 & 0.04 & 0.04 & 0.45 & \textbf{0.45} \\
        \texttt{fsmt-wmt} & 2 & 0.21 & 0.23 & 0.50 & \textbf{0.59} \\
        \texttt{fsmt-wmt} & 3 & 0.63 & 0.65 & 0.51 & \textbf{0.88} \\
        \texttt{fsmt-wmt} & 4 & 0.73 & 0.76 & 0.53 & \textbf{0.93} \\
        \texttt{fsmt-wmt} & 5 & 0.75 & 0.79 & 0.54 & \textbf{0.95} \\
        \texttt{mbart-wmt} & 1.5 & 0.07 & 0.10 & 0.38 & \textbf{0.47} \\
        \texttt{mbart-wmt} & 2 & 0.28 & 0.30 & 0.36 & \textbf{0.59} \\
        \texttt{mbart-wmt} & 3 & 0.71 & 0.73 & 0.35 & \textbf{0.91} \\
        \texttt{mbart-wmt} & 4 & 0.76 & 0.78 & 0.36 & \textbf{0.96} \\
        \texttt{mbart-wmt} & 5 & 0.78 & 0.80 & 0.35 & \textbf{0.98} \\
        \texttt{rdinet-cifar} & 1.5 & \textbf{0.61} & 0.61 & 0.48 & 0.49 \\
        \texttt{rdinet-cifar} & 2 & 0.98 & \textbf{0.99} & 0.47 & 0.84 \\
        \texttt{rdinet-cifar} & 3 & 0.99 & 1.00 & 0.48 & \textbf{1.00} \\
        \texttt{rdinet-cifar} & 4 & 0.99 & 1.00 & 0.48 & \textbf{1.00} \\
        \texttt{rdinet-cifar} & 5 & 0.99 & 1.00 & 0.48 & \textbf{1.00} \\
        \texttt{skipnet-imagenet} & 1.5 & 0.00 & 0.00 & 0.05 & \textbf{0.24} \\
        \texttt{skipnet-imagenet} & 2 & 0.00 & 0.03 & 0.14 & \textbf{0.62} \\
        \texttt{skipnet-imagenet} & 3 & 0.00 & 0.12 & 0.21 & \textbf{0.92} \\
        \texttt{skipnet-imagenet} & 4 & 0.00 & 0.31 & 0.12 & \textbf{0.95} \\
        \texttt{skipnet-imagenet} & 5 & 0.00 & 0.48 & 0.16 & \textbf{0.89} \\
\bottomrule
\end{tabular}

    \end{small}
\end{table}

\section{Generalization to Piece-wise Step Cost Functions}%
\label{sec:piece-wise-step}

We can extend our calculation for \autoref{eq:pt} from single-step SLO cost function to multiple-step cost functions.
For example, consider a multiple-step cost function with three deadlines $d_1$, $d_2$ and $d_3$, and corresponding costs $c_1$, $c_2$, $c_3$.
Such a cost function is actually decomposable into the sum of three single-step cost functions: deadline $d_1$ with cost $c1$, deadline $d_2$ with cost $c_2 - c_1$, and deadline $d_3$ with cost $c_3 - c_2$.
Therefore, we can compute the priority score for each of the single-step cost function and sum up the results to get the priority score for the multiple-step cost function.

\end{document}